\documentclass[conference,compsoc]{IEEEtran}

\usepackage{graphicx}
\usepackage{xcolor}
\usepackage{tabularx}
\usepackage{colortbl}
\usepackage{booktabs}
\usepackage[hidelinks]{hyperref}
\usepackage{msc}
\usepackage{tikz}
\usetikzlibrary{arrows.meta, positioning, fit}

\usepackage{orcidlink}

\usepackage{xparse}   % http://ctan.org/pkg/xparse
\NewDocumentCommand{\rot}{O{45} O{1em} m}{\makebox[#2][l]{\rotatebox{#1}{#3}}}%

\ifCLASSOPTIONcompsoc
  \usepackage[nocompress]{cite}
\else
  \usepackage{cite}
\fi

\ifCLASSINFOpdf
\else
\fi
\begin{document}
\title{\textsc{Ssh}-Passkeys: Leveraging Web Authentication for Passwordless \textsc{ssh}}

\author{
\IEEEauthorblockN{Moe Kayali}
\IEEEauthorblockA{University of Washington\\\texttt{kayali@cs.washington.edu}}
\and
\IEEEauthorblockN{Jonas Schmitt} 
\IEEEauthorblockA{CISPA\\\texttt{jonas.schmitt@cispa.de}}
\and
\IEEEauthorblockN{Franziska Roesner}
\IEEEauthorblockA{University of Washington\\\texttt{franzi@cs.washington.edu}}
}


\maketitle
\thispagestyle{plain}
\pagestyle{plain}

\begin{abstract}
We propose a method for using Web Authentication \textsc{api}s for \textsc{ssh} authentication, enabling passwordless remote server login with \textit{passkeys}. These are credentials that are managed throughout the key lifecycle by an authenticator on behalf of the user and offer strong security guarantees.

Passwords remain the dominant mode of \textsc{ssh} authentication, despite their well known flaws such as phishing and reuse. \textsc{Ssh}'s custom key-based authentication protocol can alleviate these issues but remains vulnerable to key theft. Additionally, it has poor usability, with even knowledgeable users leaking key material and failing to verify fingerprints. Hence, effective key management remains a critical open area in \textsc{ssh} security. In contrast, WebAuthn is a modern authentication standard designed to replace passwords, managing keys on behalf of the user. As a web \textsc{api}, this standard cannot integrate with \textsc{ssh} directly.

We propose a framework to integrate WebAuthn with \textsc{ssh} servers, by using \textsc{unix} pluggable authentication modules (\textsc{pam}). Our approach is backwards-compatible, supports stock \textsc{ssh} servers and requires no new software client-side. It offers protection for cryptographic material at rest, resistance to key leaks, phishing protection, privacy protection and attestation capability. None of these properties are offered by passwords nor traditional \textsc{ssh} keys. We validate these advantages with a structured, 
conceptual security analysis.

We develop a prototype implementation and conduct a user study to quantify the security advantages of our proposal, testing our prototype with 40 \textsc{ssh} users. The study confirms the security problems of \textsc{ssh}-keys, including 20\% of the cohort leaking their private keys. Our \textsc{ssh}-passkeys effectively address these problems: we find a 90\% reduction in critical security errors, while reducing authentication time by $4\times$ on average.

To the best of our knowledge, this work is the first to propose this integration.
\end{abstract}


% \fi
\IEEEpeerreviewmaketitle

\section{Introduction}
Secure Shell (\textsc{ssh}) is a protocol for secure remote login and network services over an insecure network~\cite{ylonen-06-ssh-prot}. Today, it is the primary remote administration mechanism for most servers and network devices~\cite{ylonen-15-sec-interactive}.  Unsurprisingly, this makes \textsc{ssh} a high-value target for adversaries. One study reported internet-facing servers face an average 250~000 unauthorized access attempts per month~\cite{alata-06-lessons-learned}.

Despite this, most \textsc{ssh} servers require only a password to access---a weak authentication mechanism with well-known flaws. Only 35\% of servers exposed to the internet require an authentication method beyond passwords~\cite{andrews-20-prevalence-pass}. While key-based authentication methods that make servers impervious to password-guessing attacks are widely available, they have seen remarkably little uptake. Key management remains an unsolved problem and is identified by the authors of \textsc{ssh} as critical open problem~\cite{ylonen-19-challenges}.

 In response to parallel issues in the internet environment, the web community has sought an alternative to passwords. The recently standardized \textit{Web Authentication} protocol~\cite{w3c-21-webauthn} allows for a complete replacement of passwords, called \textit{passkeys}. Passkeys are a new standard of credential, managed on behalf of the user by their platform. They require no passwords and offer strong security guarantees for key strength and privacy, while protecting against phishing, key theft, and reuse. Today, passkeys are supported by all major browsers and operating systems, covering 96\% of global internet users at the time of writing~\cite{caniuse-25-webauthn}.

We introduce \textsc{Ssh}-Passkeys, a method for leveraging the Web Authentication protocol for \textsc{ssh} authentication, replacing passwords with passkeys. We design a framework to integrate passkeys with \textsc{ssh} servers while maintaining backwards compatibility with current clients.

We conduct a conceptual analysis to compare the security, deployability and usability of our approach with 36 other schemes. We use the standard framework of Bonneau~\textit{et al.}~\cite{bonneau-12-the-quest} to consider 25 such factors across those three categories.

We then experimentally validate these security and usability advantages in a randomized, controlled user study. We compare \textsc{ssh}-passkeys with the baseline of traditional \textsc{ssh} keypairs. We find that passkeys reduce security errors by 83\% and critical security errors by 95\% compared to the baseline. Overall, a much larger proportion of \textsc{ssh}-passkey users are able to complete the task compared to baseline.

\subsubsection*{Contributions} Our key contributions:

\begin{itemize}
    \item We propose a method for integrating the Web Authentication \textsc{api} for \textsc{ssh} authentication in Section~\ref{sec:architecture} and describe a prototype implementation in Section~\ref{sec:implementation};
    \item We conduct a conceptual analysis of our approach against 36 alternatives in Section~\ref{sec:conceptual};
    \item We quantitatively and qualitatively verify the advantages of our approach using a user study experiment in Section~\ref{sec:evaluation}, evaluating the improvement in security analysis and finding a 95\% reduction in critical errors in Section~\ref{sec:evaluation:security}.
    \item We also conduct a usability analysis based on the user study, finding significantly improved usability and a 50\% higher score on the system usability scale, in Section~\ref{sec:evaluation:usability}.
\end{itemize}

\section{Background and Motivation}

\subsubsection*{Background} As a secure replacement for earlier protocols like \texttt{telnet}, \textsc{ssh} cryptographically encrypts and authenticates sessions. User authentication can be performed with asymmetric cryptography or with pre-shared secrets.\footnote{Server authentication is always done asymmetrically.} Standard \textsc{ssh} implementations---such as Open\textsc{ssh}---delegate pre-shared secret authentication to the operating system, while using their own custom implementation for key-based authentication.

SSH utilizes its own format for public-private key pairs~\cite{galbraith-06-ssh-key-file}. It is up to the user to manage the key pairs throughout their lifecycle and secure the on-disk representations. This is primarily because \textsc{ssh} was developed to work without any centralized key infrastructure~\cite{ylonen-19-challenges}. Key management is notoriously difficult. Past studies have been unanimous in finding that offloading key management to users leads to user frustration and critical security mistakes~\cite{lerner-17-confidante, whitten-99-johnny-encrypt}.

Standardized cryptography \textsc{api}s such as \textsc{pkcs} \#11~\cite{oasis-22-pkcs} and \texttt{X.509}~\cite{igoe-11-cert-ssh} exist and are supported out-of-the-box by \textsc{ssh} servers. However, these interfaces were designed with enterprise use cases in mind (e.g., certificate authorities) and so they have seen even less adoption than \textsc{ssh} public-key authentication.

\textit{Passkeys} are single-factor, passwordless credentials. This is the common name for \textsc{fido2} credentials managed through the Web Authentication \textsc{api}. These are asymmetric cryptography public-private keypairs that are managed end-to-end on behalf of the user by a platform authenticator (commonly their operating system) or optionally a roaming authenticator (commonly a \textsc{fido2} keyfob).  A unique, strong key is generated for each account. They have been positioned as the ``password-killer.'' They are an evolution of the earlier Universal Second Factor (\textsc{u2f}) and \textsc{fido} standards, which primarily related to hardware keys as second-factor replacements to text messages and time-based codes (\textsc{totp}).

Open\textsc{ssh} supports \textsc{u2f} and \textsc{fido} keys as of October 2020~\cite{openSSH-20-release}. It also has partial support for \textsc{fido}2 features such as \textit{resident keys}.  However, this support is only as a key storage protocol and offers no key management features. Keys are still generated manually with \texttt{ssh-keygen}. It does not support platform authenticators, the built-in \textsc{os} authenticators used by most clients. In short, \textsc{ssh} does not currently have passkey support.

\subsubsection*{Motivation} Key management failures are a serious concern: an \textsc{ssh} key-theft attack against Sony Pictures~\cite{kovacs-14-reactions-sony} in 2014 cost \$41 million in damages~\cite[p. 54]{sony-14-20f}, while another attack also related to key materials in 2011 leaked the financial and personal information of 77 million users. Not only are traditional servers vulnerable: in one vehicle telematics unit, unauthorized remote control of car throttle and brakes is possible because of poor \textsc{ssh} key management~\cite{foster-15-fast-and-vuln}. 

Vulnerability is common: one North American bank reports several \textit{million} \textsc{ssh} keys in circulation internally, while another tech company reports 30~000 machines with a single shared host key~\cite{ylonen-19-challenges}. Privileged access management system generally do not control SSH keys ~\cite[Ch. 4.4]{ylonen-15-sec-interactive}---90\% of organizations do not have an inventory of SSH keys.

Meanwhile, passkeys show promise as a solution. Passkeys are by design immune to a host of common security problems, such as weak keys, key reuse and phishing. The security advantages of similar approaches such as \textsc{u2f}, a precursor of passkeys, are well understood in the literature: studies report good overall usability, with $2\times$ reduction in login times for average users~\cite{lang-16-practical-scnd-factor}.

Several challenges prevent the integration of WebAuthn \textsc{api}s with \textsc{ssh}. First, WebAuthn is a relatively new standard: it was standardized in 2019 while \textsc{ssh} is a well-established protocol from 1999. While in 2025 passkey support is nearly universal with 96\% of online users, it was only added to Chrome and Safari browsers in the year 2022. Second, the security assumptions of the web security communities and the server administrator committees differ: this can be seen in the very different security model of \textsc{ssh} host keys and \textsc{tls} certificates. Generally, web technologies rely on a public-key infrastructure and canonical domain names while \textsc{ssh} can make neither of those assumptions. Finally, the \textsc{ssh} protocol is updated at a much more conservative frequency than web protocols. Backwards compatibility is paramount as many \textsc{ssh} installations in critical infrastructure cannot be readily updated.

\begin{figure}
    \centering
    \includegraphics[width=\linewidth]{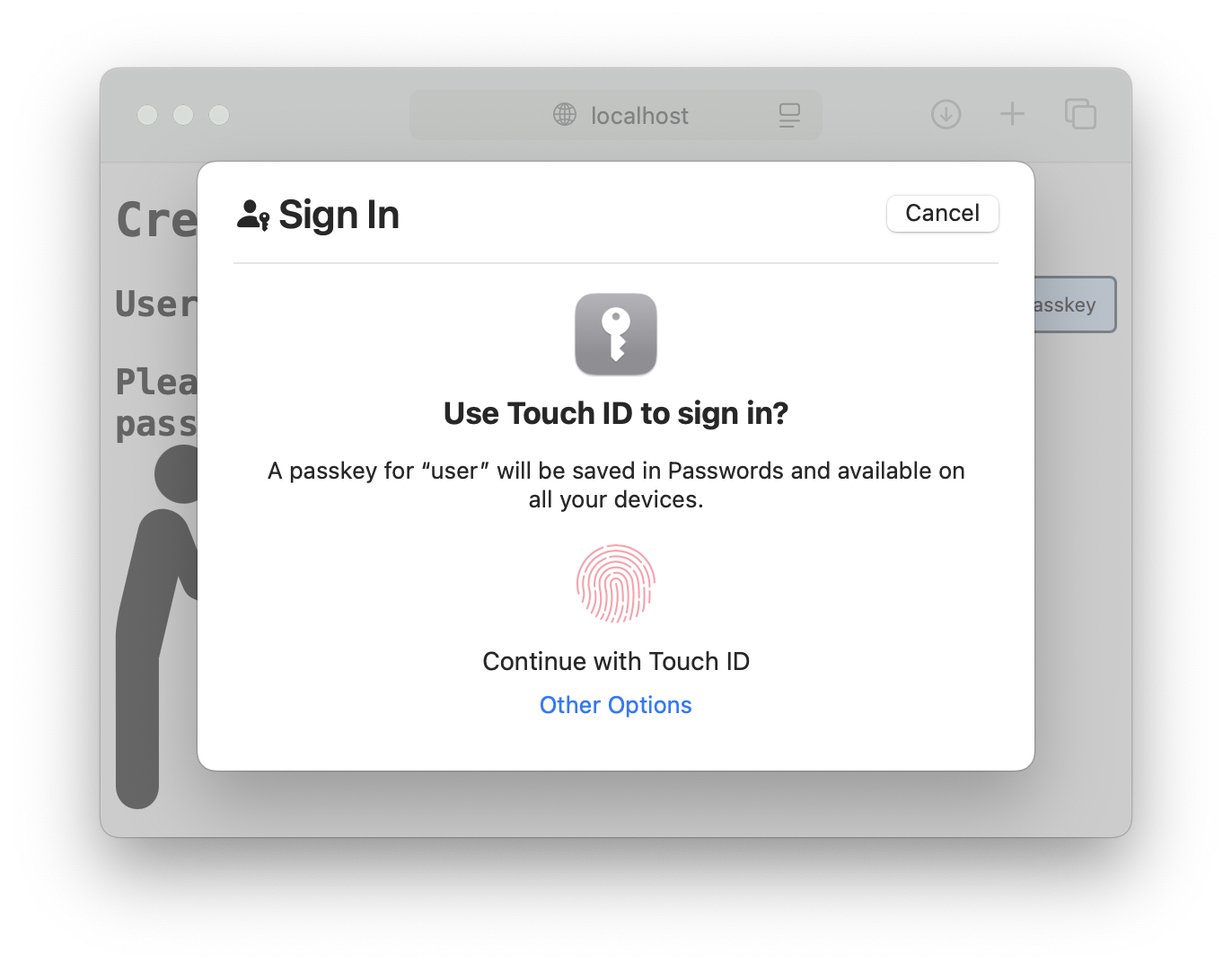}
    \caption{Screenshot of \textsc{ssh}-passkey interface. The user is registering a passkey for a first time login to an \textsc{ssh} server with the \textsc{ssh}-passkey module installed. On this platform, the passkey is secured with fingerprint access by default.}
    \label{fig:screenshot}
\end{figure}

\section{Approach}
\label{sec:approach}

Today, the typical \textsc{ssh} has two choices: login with a password or with \textsc{ssh} public key authentication. Passwords are simple but vulnerable to well-known security risks. The more security conscious user chooses keypair. She first must generate a keypair manually (e.g. using \texttt{ssh-keygen}), selecting her choice of crytographic scheme and key strength, and secure it with a passphrase. If she doesn't use a passphrase, then the key material, stored on disk, will be readily vulnerable to theft. Then, the user must find a method of transferring the public key to the server she wants to access. This may be via scripts like \texttt{ssh-copy-id} or a manual transfer with a tool like \texttt{scp}.

We envision a system where the user instead is provided a one-time \textsc{url}. Then he visits that link, he is presented with an interface as seen in Figure~\ref{fig:screenshot}. He provides explicit consent by providing his device's fingerprint or unlock pin. Then, all the following is performing on his behalf: a strong key, unique to that server is generated on a hardware security module; the identity of the server is stored and tied to the generated key; and the key is transferred to the server, where it is added to the authorized list of keys for that user. After this registration, when the user attempts to log-in via \textsc{ssh}, he is presented with a \textsc{url}, he navigates to that \textsc{url}. His browser verifies the server on his behalf and authenticates him with his private key. Back in the terminal, the user finds himself logged in with no password or passphrase.

We now show the approach followed to realize this vision. We prioritize compatibility with the \textsc{ssh} protocol, server and client. For this reason, we develop an architecture that does not require patching the ssh server nor client.

With these goals in mind, we opt to build an approach utilizing pluggable-authentication modules. Pluggable authentication modules (\textsc{pam}) allow for integrating different authentication schemes with an operating system by exposing a uniform underlying \textsc{api}~\cite{samar-95-pam-unified}. They are the standard user authentication approach in Unix systems. In Linux, custom \textsc{pam} modules can be developed with the \texttt{libpam} library. Then a server like Open\textsc{ssh} can be configured to use \textsc{pam} for user authentication.

Figure~\ref{fig:system-diagram} provides an overview of the architecture.

\subsection{Architecture}
\label{sec:architecture}

\begin{figure}
    \centering
\begin{tikzpicture}[
    font=\sffamily\scriptsize,
    sblock/.style={
        draw, thick, fill=green!5,
        rounded corners=1pt, minimum width=2.3cm, minimum height=0.65cm
    },
    cblock/.style={
        draw, thick, fill=blue!5,
        rounded corners=1pt, minimum width=2.3cm, minimum height=0.65cm
    },
    srect/.style={
        draw, thick, green!50!black, rounded corners=2pt, inner ysep=2pt, inner xsep=3pt, fit=#1
    },
    crect/.style={
        draw, thick, blue!80, rounded corners=2pt, inner ysep=2pt, inner xsep=3pt, fit=#1
    },
    arr/.style={
        <->, >=latex
    },
    narr/.style={
        arr, red!85
    }
]

%%%% Left: Server-side blocks (stacked)
\node[sblock] (ssh) {OpenSSH Server};
\node[sblock, below=10pt of ssh] (pam) {pam.d daemon};
\node[sblock, below=10pt of pam, minimum width=3.3cm, minimum height=1.5cm, anchor=north] (pammod) {};

% Place the text "Custom PAM Module" at the top of pammod
\node[anchor=north west] at ([xshift=4pt,yshift=-4pt]pammod.north west) {\textsc{Ssh}-passkey \textsc{Pam} Module};

% Now place the web node inside pammod, e.g. centered
\node[sblock, right=6pt of pammod.west, yshift=-5pt, anchor=west] (web) {Challenge Webserver};

%%%% Right: Client-side blocks (stacked)
\node[cblock, right=2cm of ssh] (term) {Terminal};
\node[cblock, below=10pt of term] (browser) {Modern Browser};
\node[cblock, below=10pt of browser] (oswebauth) {OS WebAuthn};
\node[cblock, below=10pt of oswebauth] (tpm) {TPM};

%%%% Fit rectangles (server/client)
\node[srect=(ssh)(pam)(pammod)(web), label=below:{\scriptsize\textbf{Server-side}}] {};
\node[crect=(term)(browser)(oswebauth)(tpm), label=below:{\scriptsize\textbf{Client-side}}] {};

%%%% Server-side arrows
\draw[arr] (ssh) -- (pam);
\draw[arr] (pam) -- (pammod);

%%%% Client-side arrows
\draw[arr] (term) -- (browser);
\draw[arr] (browser) -- (oswebauth);
\draw[arr] (oswebauth) -- (tpm);

%%%% Network connections 
\draw[narr] (ssh.east) -- (term.west)
    node[midway, above, inner sep=2pt] {\quad \quad Network};
\draw[narr] (web.east) ++(0,0.18) -- ++(1.2,0)
    |- ([xshift=6pt]browser.west);

\end{tikzpicture}
\caption{System diagram of our framework for passkey based \textsc{ssh} authentication. The \textsc{Ssh}-passkey module is implemented by this work.}
    \label{fig:system-diagram}
\end{figure}
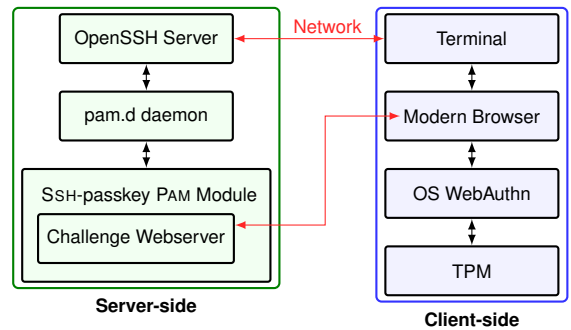

This agnostic approach can be applied to any Unix-based operating system. Our \textsc{pam} module is installed on the server by the administrator. The \textsc{pam} module is registered with the operating system and the \textsc{ssh} daemon configuration is updated to use it. When the \textsc{ssh} server calls the \textsc{pam} bindings to authenticate a user, the operating system spins up our \textsc{pam} module. The module then initiates the challenge webserver. These parts can be seen in Figure~\ref{fig:system-diagram}.

The left half of Figure~\ref{fig:system-diagram} shows the server-side components in green, with the client-side parts shown in blue on the right. Server side, we require the standard components of an \textsc{ssh} server and a \textsc{pam} daemon. The additional code required is our module, the \textsc{ssh}-passkey \textsc{pam} module. This module ships with a built-in webserver, used to host the WebAuthn protocol and serve the client browser with the libraries they need.

On the client side, only the standard component are involved: the terminal, browser with WebAuthn support, and platform authenticator (such as \textsc{os} with WebAuthn support and a trusted platform module \textsc{tpm}). The control flow moves between the terminal and the browser then the user clicks on the authentication or registration \textsc{url} provided to them by our \textsc{pam} module through the \textsc{ssh} protocol.

This means that the only new software that need to be installed is the custom PAM module on the server side.
 
\begin{figure}
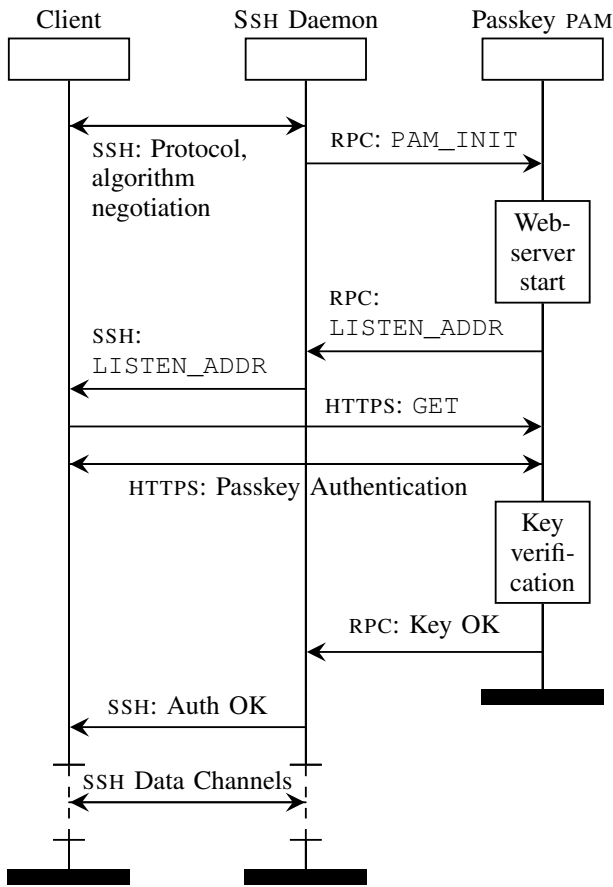

    \centering
\begin{msc}[msc keyword=Protocol, draw frame=none, environment distance=0in, instance distance=0.6in]{Authentication with \textsc{ssh}-passkeys}
  \declinst{A}{Client}{}
  \declinst{B}{\textsc{Ssh} Daemon}{}
  \declinst{C}{Passkey \textsc{pam}}{}
  \mess[text width=1in, label position=below, arrow scale=3.0]{\textsc{ssh}: Protocol, algorithm negotiation}{B}{A}
  \mess[arrow scale=3.0]{}{A}{B}
  \nextlevel
  \mess[arrow scale=3.0]{\textsc{rpc}: \texttt{PAM\_INIT}}{B}{C}
  \nextlevel
  \action{Web-server start}{C}
  \nextlevel[4]
  \mess[text width=1in, arrow scale=3.0]{\textsc{rpc}: \texttt{LISTEN\_ADDR}}{C}{B}
  \nextlevel
  \mess[text width=1in, arrow scale=3.0]{\textsc{ssh}: \texttt{LISTEN\_ADDR}}{B}{A}
  \nextlevel
  \mess[text width=1in, label position=above right, arrow scale=3.0]{~~\textsc{https}: \texttt{GET}}{A}{C}
  \nextlevel
  \mess[label position=below, arrow scale=3.0]{\hspace{-0.0825in}\textsc{https}: Passkey Authentication}{A}{C}
  \mess[arrow scale=3.0]{}{C}{A}
  \nextlevel
  \action{Key verification}{C}
  \nextlevel[4]
  \mess[arrow scale=3.0]{\textsc{rpc}: Key OK}{C}{B}
  \nextlevel
  \stop*{C}
  \nextlevel
  \mess[arrow scale=3.0]{\textsc{ssh}: Auth OK}{B}{A}
  \nextlevel
  \regionstart{coregion}{A}
  \regionstart{coregion}{B}
  \nextlevel
  \mess[arrow scale=3.0]{\textsc{ssh} Data Channels}{B}{A}
  \mess[arrow scale=3.0]{}{A}{B}
  \nextlevel
  \regionend{A}
  \regionend{B}
\end{msc}
\caption{\textit{Protocol diagram: message sequence chart for successful login with SSH-passkeys.} The client connects to the \textsc{ssh} server and begins standard protocol negotiation. The \textsc{ssh} daemon communicates with the \textsc{pam} authentication module and receives a listening \textsc{http} address. This address is forwarded to the client. The client completes WebAuthn authentication with webserver hosted by the \textsc{pam}. \textsc{Pam} then returns a success message to \textsc{ssh} server, which allows login to continue.}
    \label{fig:protocol-diagram}
\end{figure}

\subsection{Implementation}
\label{sec:implementation}

We develop a prototype implementation targeting the Linux kernel and the Open\textsc{ssh} server. We target this platform because it is dominant, by far, for internet-facing servers and cloud hosts. In principle, the \textsc{pam} interface is universal to \textsc{unix}-like operating systems and is supported in other \textsc{unix} flavors such as \textsc{bsd} or MacOS. Figure~\ref{fig:protocol-diagram} illustrates the protocol we describe now.

On the server-side, our module can be installed as a package. This consists of \texttt{C} code written to target the \texttt{libpam} library. Once installed via a configuration update in \texttt{/etc/pam.d}, the \textsc{ssh} server has \textsc{pam} support enabled by setting the \texttt{UsePAM} flag in \texttt{sshd\_config}.

Upon a user authentication request, the \textsc{ssh} server calls our library via the \texttt{libpam} interface. The \textsc{ssh}-passkeys \textsc{pam} module then spins up a web application using the \texttt{flask} framework. This web application uses \texttt{py-webauthn} to implement the server-side parts of the Web Authentication protocol. Once it is ready to receive requests, the server generates a unique \textsc{url} and passes this listening address back to the \textsc{ssh} server using a \texttt{PAM\_TEXT\_INFO} message. The server forwards this message to the client and the user navigates to that page in a browser. Upon user connection, our web application sends the authentication or registration page and a copy of the \texttt{simplewebauthn.js} Javascript library. This library implements the client-side parts of the Web Authentication protocol. The web application is aware of the \textsc{ssh} authentication process and so the web page served is specific to the user, including the details of their authentication session. 

The javascript library calls the browser bindings with the authentication challenge: the operating system passkey prompt appears to the user. Once the user successfully completes the OS prompt, the signed challenge is returned through the library to continue the Web Authentication protocol. Then, the \texttt{py-webauthn} library receives the challenge-response, decodes and verifies it. It accordingly returns a \texttt{PAM\_SUCCESS} or \texttt{PAM\_AUTH\_ERR} message, as appropriate, to the \textsc{ssh} server. The \textsc{pam} module, including the web application, are then terminated until the next authentication.

To facilitate registration, the \textsc{pam} library can be called to generate a one-time \textsc{url} that allows the user to create the passkeys and connect them to their account. This link would be transmitted to the user out-of-band during onboarding, in line with how public keys would be exchanged with an administrator during traditional \textsc{ssh} keypair configuration.

A screenshot illustrating the interface presented to the user is shown in Figure~\ref{fig:screenshot}. The codebase for the module can be obtained at \url{https://anonymous.4open.science/r/ssh-passkeys/}

\subsection{Analysis}
\label{sec:conceptual}

\subsubsection*{Conceptual Security Analysis} We evaluate our approach under the Bonneau \textit{et al.} framework~\cite{bonneau-12-the-quest}. This framework sets out twenty-five factors to be considered for any replacement for passwords. A full description of those factors is omitted for brevity---they can be found in the original paper~\cite{bonneau-12-the-quest}. Broadly, they occur in three categories: \textit{usability}, which is the burden imposed on users by the system; \textit{deployability}, or the burden imposed on system administrators and developers by the system; and \textit{security}, which are the integrity, confidentiality and availability guarantees provided.

\begin{table*}[t]
\setlength\tabcolsep{1pt}
\centering
\caption{Summary of our scheme under the Bonneau et al. framework. We compare with 36 methods. A full dot indicates support, a hollow dot partial support and an empty space indicates no support. Green shades factors where we exceed the baseline, red where we do not.}
\begin{tabularx}{0.85\textwidth}{@{\hskip 0.25in}lcccccccc@{\hskip 0.25in}cccccc@{\hskip 0.25in}ccccccccccc@{\hskip 0.25in}}
\toprule
\multicolumn{1}{c}{} & \multicolumn{8}{c}{\textbf{Usability}}
                      & \multicolumn{6}{c}{\textbf{Deployability}}
                      & \multicolumn{11}{c}{\textbf{Security}} \\

\multicolumn{1}{c}{}
    & \rot{Memorywise Effortless}
    & \rot{Scalability for Users}
    & \rot{Nothing to Carry}
    & \rot{Physically Effortless}
    & \rot{Easy to Learn}
    & \rot{Efficient to Use}
    & \rot{Infrequent Errors}
    & \rot{Easy Recovery from Loss}
    & \rot{Accessible}
    & \rot{Negligible Cost per User}
    & \rot{Server Compatible}
    & \rot{Client Compatible}
    & \rot{Mature}
    & \rot{Non-Proprietary}
    & \rot{Physical Observation}
    & \rot{Targeted Impersonation}
    & \rot{Throttled Guessing}
    & \rot{Unthrottled Guessing}
    & \rot{Internal Observation}
    & \rot{Leaks from Other Verifiers}
    & \rot{Phishing}
    & \rot{Theft}
    & \rot{No Trusted Third-Party}
    & \rot{Explicit Consent}
    & \rot{Unlinkable}\\
\midrule

\textbf{\textsc{Ssh}-Passkeys} (this work) & \cellcolor{green!30}$\bullet$ & \cellcolor{green!30}$\bullet$ & \cellcolor{red!30}$\circ$ & $\bullet$ & \cellcolor{green!30}$\bullet$ & \cellcolor{green!30}$\bullet$ & $\bullet$ & & $\bullet$ & $\bullet$ & \cellcolor{red!30} & \cellcolor{green!30}$\bullet$ & $\bullet$& $\bullet$ & $\bullet$ & $\bullet$ & $\bullet$ & $\bullet$ & \cellcolor{green!30}$\bullet$ & $\bullet$ & \cellcolor{green!30}$\bullet$ & \cellcolor{green!30}$\bullet$ & $\bullet$ & $\bullet$ & \cellcolor{green!30}$\bullet$ \\

\textbf{Passwords} & & & $\bullet$ & & $\bullet$ & $\bullet$ & $\circ$ & $\bullet$ & $\bullet$ & $\bullet$ & $\bullet$ & $\bullet$ & $\bullet$& $\bullet$ & & $\circ$ & & & & & & $\bullet$ & $\bullet$ & $\bullet$ & $\bullet$ \\

\textbf{SSH Keys} (baseline) & $\circ$ & $ $ & $\bullet$ & $\bullet$ &  & $\circ$ & $\bullet$ & & $\bullet$ & $\bullet$ & $\bullet$ & & $\bullet$ & $\bullet$ & $\bullet$ & $\bullet$ & $\bullet$ & $\bullet$ &  & $\bullet$ & & & $\bullet$ & $\bullet$ & $ $ \\ \midrule

\textbf{\# Schemes Worse} & $27$ & $21$ & $10$ & $34$ & $6$ & $27$ & $26$ & $0$ & $20$ & $18$ & $0$ & $10$ & $18$ & $17$ & $23$ & $19$ & $20$ & $22$ & $27$ & $15$ & $12$ & $9$ & $12$ & $3$ & $9$ \\
\textbf{\# Schemes Tied} & $8$ & $14$ & $7$ & $1$ & $29$ & $8$ & $9$ & $16$ & $15$ & $17$ & $29$ & $25$ & $17$ & $18$ & $12$ & $16$ & $15$ & $13$ & $8$ & $20$ & $23$ & $27$ & $23$ & $32$ & $26$\\
\textbf{\# Schemes Better} & $0$ & $0$ & $18$ & $0$ & $0$ & $0$ & $0$ & $19$ & $0$ & $0$ & $6$ & $0$ & $0$ & $0$ & $0$ & $0$ & $0$ & $0$ & $0$ & $0$ & $0$ & $0$ & $0$ & $0$ & $0$\\
\bottomrule
\end{tabularx}
\label{tab:comparision}
\end{table*}

We compare with 35 authentication methods noted by prior work~\cite{bonneau-12-the-quest}, as well as an additional one relevant to this setting, \textsc{ssh} key-based authentication---to make a total of 36 comparison methods. Table~\ref{tab:comparision} shows the summary of our evaluation.

Overall, we find that our approach fulfills 22 and a half factors out of 25 possible factors. This compares favorably with other approaches. We now break down the scoring in detail across the three categories.

\subsubsection*{(1) Usability} Usability ranks in the 89\% percentile among all other methods. We offer 6.5 out of 8 possible benefits. Advantages over passwords include nothing to memorize, usage effort not scaling with number of credentials. It is also physically effortless, defined in this framework as requiring effort beyond pressing a button. We also claim that our approach is easy to learn, a hypothesis we will experimentally validate at length in Section~\ref{sec:evaluation:usability}. 

Our usability limitations stem from needing to carry the platform authenticator, though in many cases the platform authenticator is built-in to the client device. However, if one was attempting to login via a guest computer (say at a library), they would indeed need a phone or other authenticator as a roaming mechanism. This is an improvement over \textsc{ssh}-keys but worse than passwords, which are naturally portable.

The largest usability hurdle is recovery from loss. Loss of the platform authenticator, say in case of laptop theft, would lead to loss of login ability to all servers. As such, in the web setting, some servers require registering at least two passkeys. In case one platform is lost, the other may be used as a backup. Some platforms offer different schemes for syncing passkeys, but these may compromise some security guarantees. Conservatively, we deny our implementation this benefit factor in the ranking.

\subsubsection*{(2) Deployability} Deployability ranks in the 89\% percentile. The main downside here is \textit{server-compatible} factor, as we need to install a package on the server side. A line must also be added to the \textsc{ssh} server configuration to activate our module. While not disruptive, it is a one-time action that needs to be performed by the server administrator. Passwords and \textsc{ssh} keys naturally enjoy native support in all \textsc{ssh} servers.

We also note that depending on threat tolerance \textsc{ssh}-passkeys can be implemented in a gracefully-degrading fashion, such that users may login with legacy credentials if they do not have passkeys. Those users may be then upgraded to passkey authentication. Overall, the main usability advantage of this approach is that the client does not need any additional software. We believe this is a major practical advantage. 

Passwords and ssh-keys impose a negligible cost per user. On the other hand, passkeys require a one-time moderate cost per user: this can be a one time purchase of a \textit{roaming authenticator} or an increase in the client device bill of material (\textsc{bom}) to include hardware security modules for the \textit{platform authenticator}. While there is no additional incremental cost to the user for signing up to a new website, a one-time cost must be borne directly or indirectly by the user to obtain an authenticator.

On the remaining deployability factors, such as relying on mature specifications and not being encumbered by copyrights or patents, we are tied with passwords and \textsc{ssh} keys. All implementations, including ours, are open-source.

\subsubsection*{(3) Security} \textsc{Ssh}-passkey security factors rank in the 97\% percentile. \textsc{Ssh}-passkeys offer all the benefits in this category. This may be expected as, in the web setting, passkeys are considered very strong credentials. Passkeys succeed where passwords fail at guessing resistance, leaks and phishing. Passkeys offer advantages that traditional \textsc{ssh} keys do not have: they are resistant to internal observation, theft and linking across verifiers.

WebAuthn requires the generation of keypairs with high key entropy. Generally these are \textsc{ecdsa} with 256 bits, \textsc{e}d\textsc{dsa} with 256 bits, \textsc{rsa} with 2048 bits or more, and \textsc{sha}-256 or stronger. This is substantially stronger than the typical key strength used in passwords and comparable to \textsc{ssh} keys.

\textsc{Ssh} keys are generally linkable to their users as the typical usage is making a one keypair per device, where the one fixed public key is distributed to all servers. As such, it is straightforward for the verifiers to link the logins. Because the WebAuthn standard requires the generation of a fresh passkey for each origin, the credentials cannot be correlated across logins on different servers.

Despite being a web standard, one big advantage of passkeys is not relying on trusted third parties, as is the case with \textit{e.g.} the \textsc{tls} security model and its certificate authorities (\textsc{ca}s). The authentication standard is fully decentralized, with no 3rd party required for key creation or verification, just like \textsc{ssh} keys and passwords.

\subsubsection*{\textsc{Ooda} Loop} To quantify the security errors in the empirical portion of the study, we develop a taxonomy based on the \textsc{ooda} loop. The \textsc{ooda} loop~\cite{boyd-96-essence}---observe, orient, decide and act---was developed to analyze human decision making in contested, high-stakes environments. It originates in military doctrine but has been applied to business, aviation and cybersecurity.

In this model, any agent follows the four stages of the cycle sequentially. In the first step, \textit{Observe}, the agent collects cues from the environment. The agent decides which variables to observe and were to spend his attention. Then, in \textit{Orient}, the observations are then synthesized into a model of the situation. This includes assessing which observations are relevant and which are not. The data are interpreted and ambiguities decided on. Once the agent has built his mental model, he continues to the \textit{Decide} step: based on his model, the user considers different counterfactuals and selects the one which best achieves his goals. Finally, the agent reaches the \textit{Act} step, in which he must have the knowledge and ability to affect the desired change. One iteration of the process complete, the agent repeats the \textsc{ooda} cycle until goal completion or failure.

The two inner steps of the cycle, \textit{orient} and \textit{decide}, involve higher-level understanding of the problem: they relate to the user's mental model. The two outer steps of the loop, \textit{observe} and \textit{act}, are ``lower level'' and relate to the user's practical ability to perceive the environment accurately and execute their planned actions. As such, errors in the inner steps can be seen as more critical and harder to recover from.

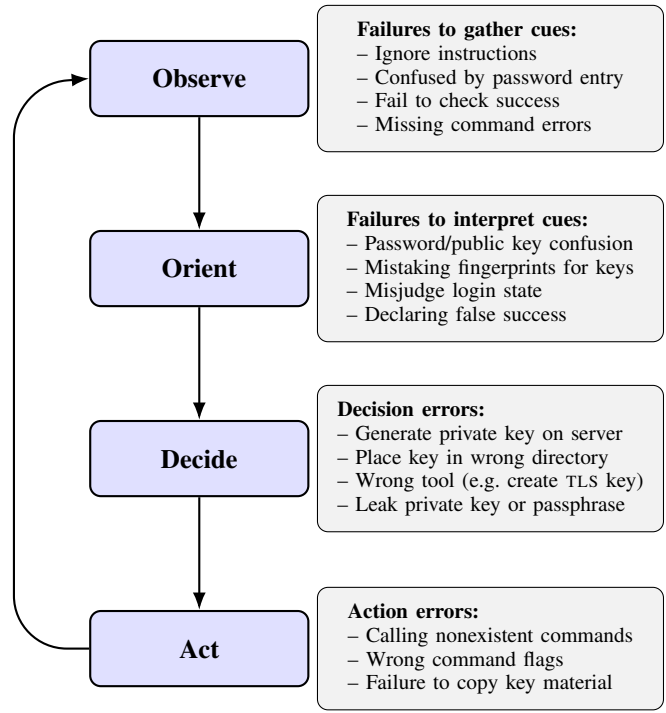
\begin{figure}
    \centering
\begin{tikzpicture}[
    font=\sffamily,
    phase/.style={
        draw,thick,rounded corners,fill=blue!12,
        minimum width=2.9cm, minimum height=1cm,
        font=\bfseries, text centered
    },
    ann/.style={
        draw,rounded corners,fill=gray!10,
        minimum width=4.6cm, font=\footnotesize,
        inner sep=6pt
    },
    every node/.style={align=left},
    >=Latex
    ]

% OODA PHASES (vertical)
\node[phase]    (obs)                                   {Observe};
\node[phase, below=1.5cm of obs]   (ori)                {Orient};
\node[phase, below=1.5cm of ori]   (dec)                {Decide};
\node[phase, below=1.5cm of dec]   (act)                {Act};

% Arrows down and looping up
\draw[thick,->] (obs) -- (ori);
\draw[thick,->] (ori) -- (dec);
\draw[thick,->] (dec) -- (act);
\draw[thick,->,rounded corners=20pt] (act.west) -- ++(-1,0) -- ++(0,7.5) -- (obs.west);

\node[ann, right=0.1cm of obs]
    (obsann) {
        \textbf{Failures to gather cues:}\\
        -- Ignore instructions\\
        -- Confused by password entry\\
        -- Fail to check success\\
        -- Missing command errors
    };

\node[ann, right=0.1cm of ori]
    (oriann) {
        \textbf{Failures to interpret cues:}\\
        -- Password/public key confusion\\
        -- Mistaking fingerprints for keys\\
        -- Misjudge login state\\
        -- Declaring false success
    };

\node[ann, right=0.1cm of dec]
    (decann) {
        \textbf{Decision errors:}\\
        -- Generate private key on server\\
        -- Place key in wrong directory\\
        -- Wrong tool (e.g. create \textsc{tls} key)\\
        -- Leak private key or passphrase
    };

\node[ann, right=0.1cm of act]
    (actann) {
        \textbf{Action errors:}\\
        -- Calling nonexistent commands\\
        -- Wrong command flags \\
        -- Failure to copy key material
    };

\end{tikzpicture}
\caption{Conceptual model of user errors in our evaluation based on the \textsc{ooda} cycle}
    \label{fig:ooda}
\end{figure}

We built the taxonomy by having two separate researchers code the errors in the transcripts of the participant actions after running the user-study experiment. The codes were then merged and regularized. This yielded a canonical list of errors with were then classified in the steps of the \textsc{ooda} cycle. A third reviewer then checked the results for uniformity and consistency. An excerpt of the resulting taxonomy is shown in Figure~\ref{fig:ooda}.

In the observe category, we include errors such as ``ignoring instructions'' where the user fails to head a clear instruction, such as being unable to find the server hostname. Common errors in the orient category include misunderstanding whether the current login is a password or key-based login. Decide errors include making the decision to create the private key on the server. Finally, action errors are those where the decision is correct but the user is unable to successfully effect the decision. For example, many participants failed in transferring the public key to the server despite multiple attempts with different file transfer tools.

After building the taxonomy, each of the terminals were classified as critical or non-critical security errors. Critical errors are those which, on their own, lead to a security breach. Non-critical security errors are those which, while unlikely to cause a compromise on their own, nonetheless reduce the security posture of the system. Leaking private key material is an example of the former; not verifying \textsc{ssh} host fingerprints an example of the latter.

An outline of the categories is shown in Figure~\ref{fig:ooda}.

\section{Evaluation}
\label{sec:evaluation}

\subsection{Methods}

\subsubsection*{Recruitment} We enrolled 40 participants, selected from the undergraduate computer-science student population at two major research universities. One half of the cohort was recruited in Germany and the other in the United States. Virtually all participants indicated familiarity with Linux and the command line. Users were briefed on the study and if participating rewarded with \$25 or €25 at the conclusion of the study. All forty participants remained in the study to completion, none dropped out.

\subsubsection*{Setup} We utilized a containerized server infrastructure to conduct the experiments. This was composed of three systems: one client environment and two \textsc{ssh} servers---a test and a control. All systems ran Ubuntu 24.04 LTS operating system. The control server run a vanilla Open\textsc{ssh} daemon, while the test servers additionally had our SSH-passkey PAM module installed. Docker and Docker Compose were used to spin up a fresh environment for each participant: this allowed us a consistent test environment for all experiment runs.  Code for turn-key replication of the experiment environment is available at: \url{https://anonymous.4open.science/r/ssh-passkeys/}

\subsubsection*{Study Design} We randomly assigned participants to the control or treatment arms of the study. To mitigate participant response bias~\cite{dell-12-yours-is-better}, we do not disclose to the participants which method was developed by the authors. We also do not disclose whether they are in the treatment or control arm. The protocol involved a pre-interview, the task itself and post-interview. The pre-interview introduces the task to the user and collected background information about the user. Depending on the assigned group, the participants were provided with one of sets of instructions in the form of an email. For the control group, an email providing a one-time password to log into the server was given, along with instructions to use that to connect and then set up key-based authentication. For the treatment group, a similar email, but with a sign-up \textsc{url} was provided. The instructions were otherwise identical. The participants were to perform the task without external help. An initial timeout of 20 minutes was set. If the participants hit the timeout or gave up on the task, we would allow them external help and another 20 minutes, for a total timeout of 40 minutes.

To prove that they had accessed the server correctly using their respective authentication methods, the participants were asked to create a dummy file on the server. The full command to create the indicator file was provided to the participants.

After the task was complete, in the post-interview we asked the participants to rate whether they succeeded at the task, as well as the question battery from the systems usability scale (\textsc{sus})~\cite{brooke-86-sus}. After the formal interview is concluded, we debrief the participants on the purpose of the study.

\subsubsection*{Ethics} This study involves human subjects. Informed consent was obtained from all participants. We proactively informed participants of legal rights, such as under \textsc{eu} data protection regulations, available to them. To protect participant privacy---and protect them from harms such as key leaks---all test infrastructure was provided to them rather than allowing the use of personal devices. After collection of study-relevant data, the environments were securely wiped. All study-relevant data was then anonymized after collection. We report aggregate statistics and never provide identifying data. The study was designed with the Menlo Report principles in mind~\cite{dhs-12-menlo}. The study was cleared by the respective institutional review boards (\textsc{irb}s) of both universities.

\subsection{Security Results}
\label{sec:evaluation:security}

To measure the security performance of the treatment and control group, we coded errors based on the \textsc{ooda} taxonomy presented in Section~\ref{sec:conceptual}. As explained, we categorize the security errors as critical or non-critical. Critical errors are can directly lead to serious security consequences, such as leaking the private key. On the other hand, non-critical security errors are those which, while definitely a determent to the integrity of the system, would not likely on their own lead to a breach. Examples include not verifying \textsc{ssh} host fingerprints or corrupting a key while copying it.

In summary, we find that among our participants, \textsc{ssh}-passkeys have a significant security advantage over traditional \textsc{ssh} keypairs, including a 83\% reduction in all security errors, a 95\% reduction in critical errors, a lower number of correlated errors, and the elimination of key leak mistakes made by 20\% of baseline participants. A large portion of the errors made by control group (47\%) are impossible by design with \textsc{ssh}-passkeys.

\begin{figure}
    \centering
    \includegraphics[width=0.9\linewidth]{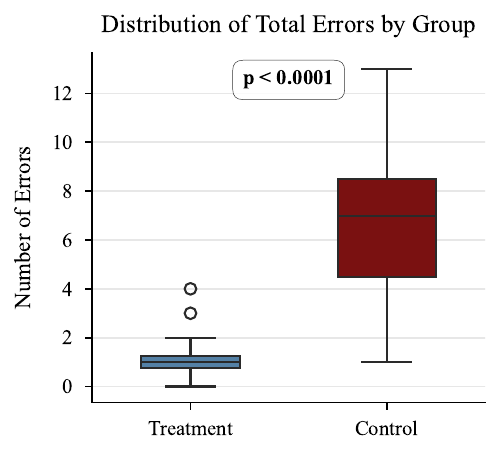}
    \caption{Distribution of Total Errors by Group}
    \label{fig:total-errors}
\end{figure}

\subsubsection*{\textsc{Ssh}-passkeys reduce security errors by 83\%} The control group, using ssh keys, made numerous security errors: average participant made 6.95 security errors. Meanwhile, the treatment group saw an average of 1.15 errors per participant, an 83\% reduction. These trends are illustrated in Figure~\ref{fig:total-errors}. Checking for statistical significance with Student's $t$-test for difference in total errors, we find that this difference is significant with $p < 0.001$.

\begin{figure}
    \centering
    \includegraphics[width=0.9\linewidth]{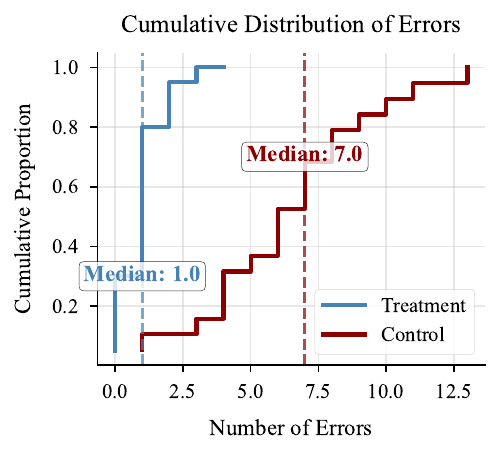}
    \caption{Cumulative Distribution of Errors}
    \label{fig:cdf-errors}
\end{figure}

Figure~\ref{fig:cdf-errors} shows the cumulative distribution of errors between the two groups. The 50\% percentile participant makes one error in the treatment group, while they makes seven errors in the control. Additionally, the treatment group is characterized by a sharp upward curve: the bulk of the population is in the low-error area. Meanwhile, the control group sees a gradual increase, reflecting a long-tail of users that make numerous errors.

\begin{figure}
    \centering
    \includegraphics[width=0.9\linewidth]{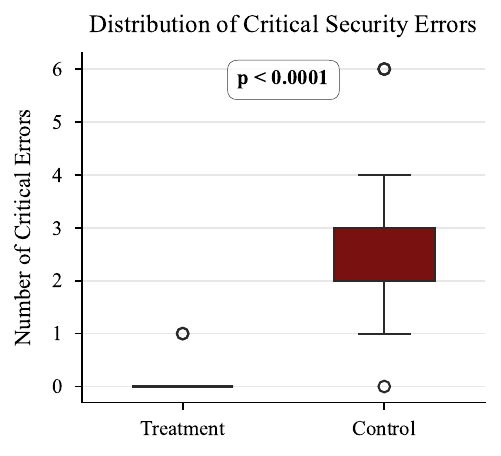}
    \caption{Average Critical Security Errors by Group}
    \label{fig:critical-errors}
\end{figure}

\subsubsection*{With \textsc{ssh}-passkeys, critical security errors are reduced by 95\%} The average user in the ssh-passkey treatment group made 0.10 critical errors, while the average user in the control group committed 2.74 such errors. This corresponds to a 95\% drop. Figure~\ref{fig:critical-errors} shows the distribution of critical errors. Applying Student's $t$-test again to check the statistical significance of the difference between the two groups yields a favorable $p$-value $p < 0.001$.

\begin{figure}
    \centering
    \includegraphics[width=0.9\linewidth]{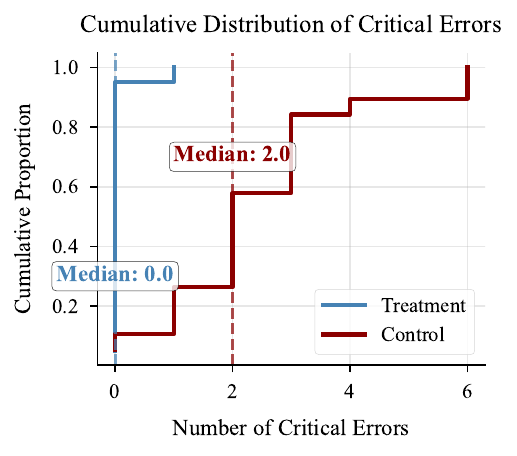}
    \caption{Cumulative Distribution of Critical Errors}
    \label{fig:cdf-critical}
\end{figure}

Figure~\ref{fig:cdf-critical} shows the cumulative distribution of critical errors. Here, the median passkey user makes no errors, while the median baseline user makes two critical errors. Similar to the distribution of non-critical errors, the baseline participant critical error counts have a large variance.

\begin{figure}
    \centering
    \includegraphics[width=0.75\linewidth]{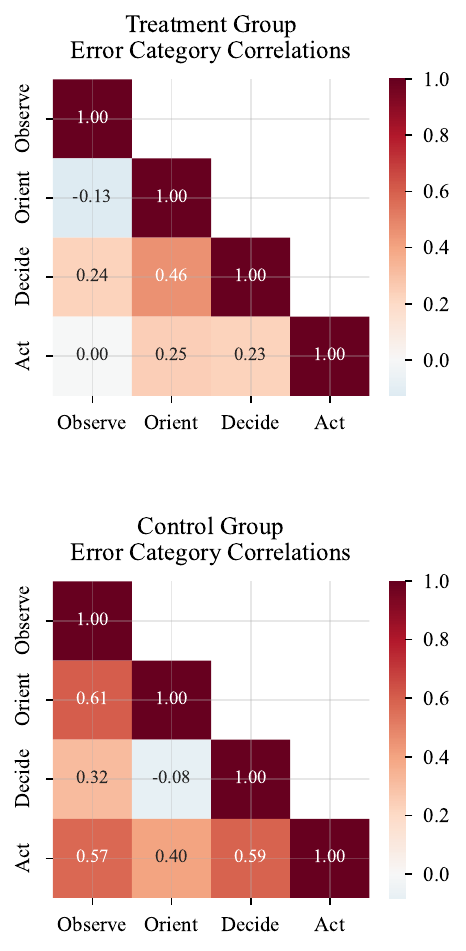}
    \caption{Correlation matrix between the error categories for the control group: if you make a mistake in one category, you’re very likely to follow-up with mistakes in others (failures cascade). Failures are well-isolated in the treatment group.}
    \label{fig:correlation-matrices}
\end{figure}

\subsubsection*{47\% of errors made by the control group are impossible, by design, with \textsc{ssh}-passkeys} When considering all the security errors made by the control group using traditional \textsc{ssh} keys, we find that 47\% of all errors made would be impossible by design in a passkey-based implementation. Of the twenty control participants, nineteen (95\%) of users made at least one such error.

\subsubsection*{\textsc{Ssh} key errors escalate, passkey errors do not} Figure~\ref{fig:correlation-matrices} shows the Pearson correlation of errors among the four \textsc{ooda} error categories. In the treatment group, we see low correlations among the error categories, indicating that errors remain isolated: making an error in one category is unlikely to lead cascading errors. The correlation among all error categories is $r=0.16$, a weak overall correlation.

Meanwhile, in the control group, errors are highly correlated. The highest correlation $r=0.61$ is between Observe-Orient errors, as might be expected. This indicates that once a user makes an error in what to observe, their mental model is likely to do off track. This is in contrast to the weak correlation ($r=0.13$) for the same pair in the treatment group. Overall, the correlation between error categories in the control group is $r=0.40$, a moderate positive correlation.

\subsubsection*{Users Frequently Leak Private Keys}

Despite a high level of conceptual understanding of public-private keypairs during the interview---where all participants were able to articulate that the private key is not for sharing---20\% of participants still indicated the private key for uploading. In all cases, it appears that the users did not understand that they were transmitting their private keys. One participant even uploaded his private key to an online document collaboration platform.

\subsubsection*{Users Rarely Secure \textsc{ssh} Keys with passphrases} Only 5 of 20 (25\%) of participants secured their \textsc{ssh} keypairs with a passphrase, the rest left their \textsc{ssh} keys unencrypted on disk. In contrast, by design all the key material of the passkey users was encrypted at rest.

\subsection{Usability Results}
\label{sec:evaluation:usability}

In addition to the wide disparity we see in the security outcomes, we consider the usability advantages of our approach. We consider a participant to be successful if they meet the following conditions: they completed the task within the time limit and did not commit any critical errors.

We categorize results that relate more to the user's ability to complete the task and their need for external help resources under \textit{usability}, while results relating to the user avoiding creating vulnerability are categorized in the previous section under \textit{security}. Naturally, these two are closely related.

In summary, we find that \textsc{ssh}-passkeys have a significant usability advantage over traditional \textsc{ssh} keypairs, including a $4\times$ speedup in workflow time, a significantly higher success rate, $2\times$ lower incidence of incorrect mental-model, a 40\% reduction in the need for external resources to complete the task, and a 50\% higher score on the system usability scale (\textsc{sus}).

\begin{table}
\caption{Task completion time in the two groups}
\centering
    \begin{tabular}{ ccc }
    \toprule
      &  Time Average (seconds) & Success rate \\ \midrule
     Control & 1306 & 2 of 20 \\
     Treatment &  \textbf{330} & \textbf{18 of 20} \\ \bottomrule     
    \end{tabular}
    \label{tab:important-stats}
\end{table}

\subsubsection*{\textsc{Ssh}-passkey workflow is $4\times$ faster}

Table~\ref{tab:important-stats} compares the completion times and success rates of the treatment and control groups. While the average baseline participant in the control group required over 21 minutes to complete the task, the average participant in the treatment group needed less than 6 minutes---a substantial speedup. 

This average completion time does not include users who failed the task, a number of whom which hit the 40 minute timeout in the control case. No users in the treatment group reached the timeout.

\subsubsection*{Passkey users complete the task; \textsc{ssh}-key users fail} The success rates show a stark difference: 18 out of 20 (90\%) passkey users succeed in the task, while only 2 out of 20 (10\%) of ssh-key users do. This can be seen in Table~\ref{tab:important-stats}. This is particularly notable in light of the test cohort being composed entirely of computer scientists. In other words, even for target populations, SSH key usability is poor but passkey usability is reasonable.

\begin{figure}
    \centering
    \includegraphics[width=0.9\linewidth]{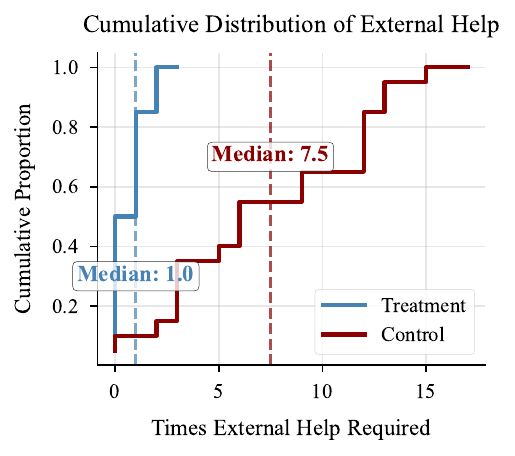}
    \caption{Cumulative distribution of external help accessed by the two groups. 45\% of passkey users accessed no external help; only 5\% of ssh-key users managed without.}
    \label{fig:external-help}
\end{figure}

Also notable is the high failure rate in the control group despite accessing external resources, shown in Figure~\ref{fig:external-help}. The median control user access external help over 7 times but these resources were not sufficient. 

\begin{figure}
    \centering
    \includegraphics[width=0.85\linewidth]{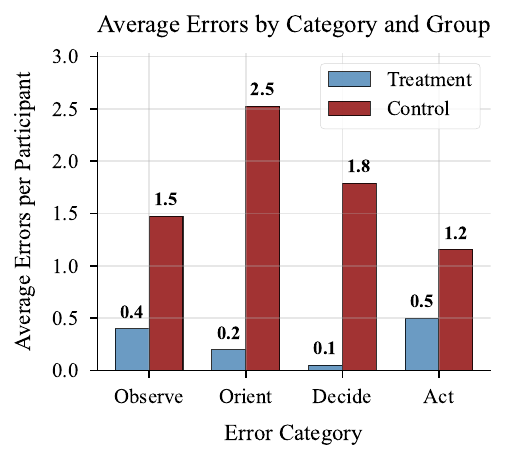}
    \caption{Errors by category. Errors are much more common in the conceptual (orient, decide) categories for the control group, representing failures of mental model. Our trend is the inverse.}
    \label{fig:errors-by-category}
\end{figure}

\subsubsection*{\textsc{Ssh} key errors are less recoverable than passkey errors}

Figure~\ref{fig:errors-by-category} shows the average number of errors the participants made, grouped by category. As discussed in Section~\ref{sec:approach}, errors in the two inner steps of the \textsc{ooda} loop (orient and decide) are more severe, as they represent mental-model deficiencies that are difficult to recover from. We see a trend where the control group users make these errors at a higher rate, for example orient-errors occur at $2.1\times$ the rate of act-errors. The inverse distribution is found in the treatment group, where errors in the two outer steps are more likely: orient-errors are $2.5\times$ \textit{less} likely than act-errors. These are more likely to represent inconsequential errors that are easily recovered from with the right mental model, such as mistyping a command.

\subsubsection*{\textsc{Ssh}-key users require external help; passkeys are self-explanatory}
Passkey users did not need to access any external assistance 45\% of the time, while only 5\% of ssh-key users did the same. In the treatment group, the average participant accessed external resources 0.80 times, while the control group needed them an average 8.05 times. The difference between the two groups is statistically significant at the $p < 0.001$ level. 

Examples of external assists include looking up \texttt{ssh-keygen} syntax or how to transfer files using \texttt{scp}. 

Figure~\ref{fig:external-help} shows the cumulative distribution for external help access between the two populations. The distribution for the passkey group is both lower in the number of required assists and more tightly clustered. Meanwhile, the control group needed a median of 7.5 assists and the 80\% percentile required 12 assists.

The most common external resource accessed by passkey users was \texttt{ssh} command syntax: 85\% of passkey keys required no external resources or only \textsc{ssh} syntax. In the control group, the external help count is high in part because of multiple lookups, where the participants try a command and it fails. For example, one participant mistyped the \texttt{scp} command and went down the wrong path trying to replace the \texttt{libcrypto} library on the client.

\begin{figure}
    \centering
    \includegraphics[width=0.9\linewidth]{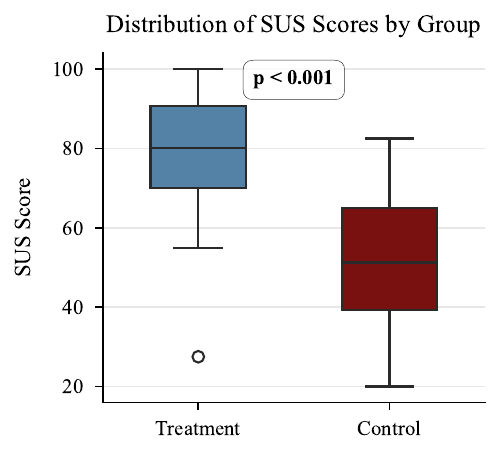}
    \caption{System usability scale (\textsc{sus}) scores. With an average score of 78.9, \textsc{ssh}-passkeys (ours) have 50\% higher usability scores than the baseline, at 52.6 average points.}
    \label{fig:sus-distribution}
\end{figure}

\subsubsection*{Passkeys have substantially higher usability score} Passkeys scored 78.9 average on the System Usability Scale (\textsc{sus}), compared to a 52.6 average for \textsc{ssh} keys. Qualitatively, these correspond to ``good'' to ``excellent'' for passkeys and ``average'' usability for the baseline. This difference is statistically significant at the $p<0.001$ lavel. Figure~\ref{fig:sus-distribution} shows the estimated distributions of \textsc{sus} scores for the two comparison populations.

\begin{figure}
    \centering
    \includegraphics[width=0.9\linewidth]{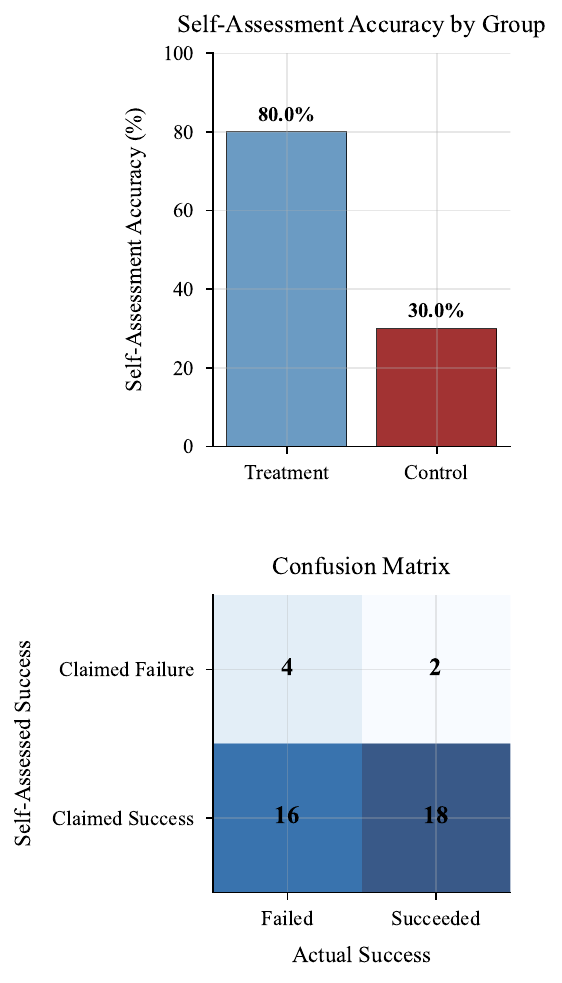}
    \caption{\textit{Self-assessment correctness}. User assessment of success is no better than chance (55\% accuracy) overall. But our users are pretty accurate. Control group is wildly overconfident. If participants claim success, they are right 53\% of the time. But if they actually succeed, they will correctly claim it 90\% of the time.}
    \label{fig:self-assessment}
\end{figure}

\subsubsection*{User self-assessment is no better than chance} Overall accuracy of the participants' self-assessment of their success compared to actual success (no critical errors, no timeout) was only 55\%, no better than chance. Figure~\ref{fig:self-assessment}(a) shows this. Passkey users were substantially more accurate in their self assessment at 80\% accuracy, compared to only 30\% of the control group. In other words, \textit{the control group was worse at identifying their own success/failure than a coin toss}.

We noted that a large number of participants were not able to correctly detect nor parse error conditions returned to them by the \textsc{ssh} key tools, for example failing to realize that a non-zero return value is an error and that they need to alter the command. These ``silent'' failures cause a false sense of confidence, as the participants believe they succeeded.

Another distinction between the two groups is \textit{calibration}. The perceptions of the passkey group are well-calibrated: when their self-assessment is wrong, it is equally likely to be overconfident (10\% of all group members) or underconfident (10\%). Meanwhile, the control group's errors are all in one direction, overconfidence (70\%) with not a single participant making an underconfidence error (0\%). 

Figure~\ref{fig:self-assessment}(b) shows the confusion matrix for self-assessment. True positives and false negatives are nearly equally prevalent (18 vs 16 participants), while true negatives are more prevalent than false negatives (4 vs 2 participants). This corresponds to a precision of $0.529$ and a recall of $0.900$. Precision is the proportion of true positives among claimed positives, while recall is the portion of positives detected correctly. Thus, these metrics can be interpreted as: if participants claim success, they are right 53\% of the time. But if they actually succeed, they will correctly claim it 90\% of the time.

\begin{figure}
    \centering
    \includegraphics[width=0.9\linewidth]{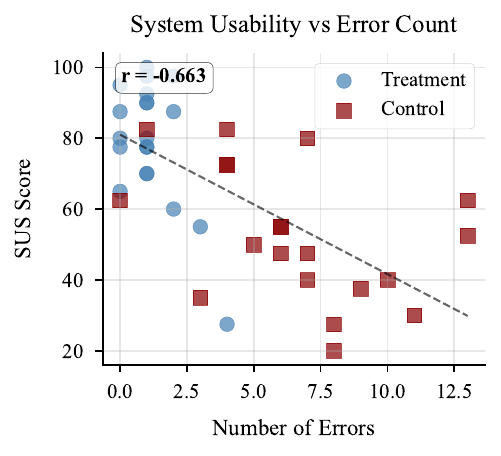}
    \caption{System usability scale (\textsc{sus}) scores vs number of errors committed, control \textit{vs} treatment. \textsc{Sus} scores are moderately correlated with the number of errors.}
    \label{fig:sus-vs-errors}
\end{figure}

\subsubsection*{Usability corresponds to security} Figure~\ref{fig:sus-vs-errors} plots the relation between \textsc{sus} scores and the number of errors per participant. The relation has a correlation of determination $R = -0.663$ and so $R^2 = 0.44$. This corresponds to a ``moderate to high'' degree of influence between the two measured variables. The negative value is indicates that as \textsc{sus} score increases, the number of expected security errors decreases.

The $R^2$ value for critical security errors is lower at $R^2 = 0.26$, interpreted as a ``weak'' predictor. Similar to the findings of the prior section, participants are not aware of critical mistakes even when filtered through the lens of the validated questions of the systems usability scale.

% \subsubsection*{Browser friction}

\section{Discussion}
\label{sec:discussion}

\subsubsection*{Many users are unable to effectively use \textsc{ssh} key authentication} Despite studying a cohort with a technical computer background, we find that only $10\%$ of participants succeed at setting up and logging-with \textsc{ssh} keypairs. The majority of participants reqiured the help of external resources and even then still commited critical security errors. An interpretation is that \textsc{ssh} key authentication is usability is not sufficient for the current cohort of developers and computer scientists.

The finding that 47\% of \textsc{ssh}-key errors in the control group are impossible by design in the passkey-based approach points to a multi-factor root cause. It is an approximately even mixture of two factors: a lower number of security benefits offered by the scheme that cause real issues in practice (47\% of errors) and poor user experience and interface (\textsc{ui}/\textsc{ux}) causing user errors that are preventable but occur predominantly in the control group (53\% of errors).

Anecdotally, instructors at our institution report that setting up SSH-keys is a major obstacle for practical assignments in computer science education, even in 400-level (4th year undergraduate) computer science courses. We developed this work in-part to help alleviate that pain point.

\subsubsection*{\textsc{ssh}-passkeys resolve well-known privacy problems} Public-key authentication in \textsc{ssh} has well-known privacy issues~\cite{roy-22-pracical-pets}. In particular, a server can learn a client's entire set of public keys regardless of whether they are registered with that server or another. Interestingly, the inverse is also true, where a user can probe a server to discover whether it accepts another user's public keys. Proof-of-concept deanonymizing \textsc{ssh} servers have been built~\cite{valsorda-20-whoami}.

\textsc{Ssh}-passkeys are immune to these issues. The WebAuthn standard requires separate keypairs be generated for each verifier. This prevents correlation across different servers. 

\subsubsection*{Web, \textsc{ssh} security models clash} The authentication infrastructure of the web and \textsc{ssh} differ. The public-key infrastructure (\textsc{pki}) model---underpinned by certificate authorities (\textsc{ca}s)---is predominantly used in \textsc{tls}. The \textit{trust-on-first-use} (\textsc{tofu}) model dominates in \textsc{ssh}: under this model, the server key material is trusted by the client on first connection and a warning is later raised if the server identity changes~\cite{wendlandt-08-perspectives}.

The impracticality of the \textsc{pki} model for server infrastructure was identified as early as the 1980s~\cite{rivest-84-expose-eave}. The differences between these security models are primarily driven by one differing core assumption: web-based technologies can assume internet access, while network protocols like \textsc{ssh} cannot. For example, the \textsc{ssh} model needs to work in air-gapped network, \textsc{tls} does not. Further, \textsc{ssh} servers may be ephemeral---while generally durable \textsc{tls} certificates are issued for domain names.

Another issue is that websites have one canonical name (the \textsc{url}) and as such, every passkey is tied to one domain. In the server administration setting, one server may have many different aliases (e.g. \texttt{worker}, \texttt{worker.local}, \texttt{worker.corp.net.}, \texttt{load-balancer.corp.net.}, \texttt{192.168.0.5})---it would be desirable to be able use any of them for \textsc{ssh} login. With the current implementation, the phishing protections of passkeys would decline to sign any request from outside the origin used for registration. The WebAuthn standard has recently introduced \textit{related origin requests} (\textsc{ror}), which allow one passkey to be released to a number of pre-registered domain aliases. This would provide a method of implementing this feature.

\subsubsection*{Broadened attack surface with \textsc{ssh}-passkeys} Our implementation of \textsc{ssh}-passkeys requires running a web-server embedded within the \textsc{pam} authentication module. While this server may be hardened since it requires only a small fraction of web protocols, this still represents a broadened attack surface for would-be attackers. Web Authentication is also a younger standard than \textsc{ssh} public-key authentication and is still evolving. While we believe the security advantages demonstrated make \textsc{ssh}-passkeys well worth it and mitigate critical and common issues, for some organizations with very high assurance requirements this increased server exposure may not be acceptable. Further, from the client's perspective connecting to an untrusted server, this exposes the client to a lot more untrusted executable code (\textit{e.g.} Javascript). While browser sandboxing is an intense area of security focus, this may still not be acceptable for the most security-sensitive of clients.

\subsubsection*{Good mental models not enough} We found that virtually all participants understood the distinction between public and private keys. Nonetheless, in practice fully 20\% attempted to upload their private key instead of their public key. One interpretation is that users know not to leak a private key but in practice they don't know what actions that translated to for \textsc{ssh} keys. Alternatively, this may be a form of social desirability bias: participants are savvy enough to understand that the experimenters expect a certain answer, so pretend to be security conscious~\cite{dell-12-yours-is-better, wash-17-self-report}. 

\subsubsection*{\textsc{Unix} user management} An advantage of the our approach building a \textsc{pam} module is that passkeys can be used beyond \textsc{ssh}. This means that our findings have broader implications for account management on \textsc{unix} systems in general. For example, \textsc{ssh}-passkey module can also be used for privilege management with \texttt{sudo}. In case of this deployment, it may be preferable to integrate with the \textsc{os} fido2 apis rather than using a web browser. See further discussion of this under \textit{Future Directions}.

\subsubsection*{\textsc{Ssh} keys are vulnerable to theft} Virtually no participants (5 out of 40) used a passphrase to encrypt their private \textsc{ssh} at rest. Since no access control is applied to the keys either, any software running as those users would be able steal their credentials. In essence, these were equivalent to storing one's password in a plaintext file. In contrast, passwords though much maligned, in the web setting, most users do not write down their passwords and only 16\% of users select passwords with unacceptable strength ($<30$ bits)~\cite{florencio-07-web-pass}.

\subsubsection*{Self-assessment is unreliable. System Usability Scale (\textsc{sus}) with caution} We find that user judgment of success is poor, being only marginally better than chance at 55\% accuracy. This is in line with prior work~\cite{wash-17-self-report}, which finds that for an overwhelming majority of tested security best practices, there is no correlation at all between user claims of adopting the measures and actual adoption. Even for a minority of measures (3 out of 13) were a correlation was statistically significant, it never exceeded a weak or moderate effect.

Similarly, we find that while the systems usability scale was moderately predictive of the number of errors, it was only poorly predicative of critical security errors. While the \textsc{sus} is well-validated for measuring usability~\cite{bangor-24-empirical-sus}, we suggest that in the security context it is predictive of user effort rather than outcomes.

\subsubsection*{Friction introduced} There is a trade-off between client compatibility and end-user usability. In particular, to achieve client compatibility, our current implementation redirects the user briefly to the browser to invoke the \textsc{os} authentication hooks. If installing software on the client side is acceptable, it would be possible to invoke those \textsc{os} calls directly from the terminal. This would remove the need to context-switch into the browser to complete the login. This may also reduce the attack surface mentioned in the earlier section.

\subsubsection*{Key Agents and Forwarding} One subtle advantage of our passkey based approach is attenuating the need for \textsc{ssh} key forwarding. Sometimes a user needs to chain \textsc{ssh} connections, for example in the common enterprise case where security policy means internal hosts can only be accessed via a bastion host. In this case, the user must configure \textit{key forwarding}, where the intermediate host has access to all the user's key material. This presents a strong security vulnerability and must not be enabled for untrusted servers. Meanwhile, with \textsc{ssh}-passkeys, the user will simply visit the \textsc{url} provided by the end host and decide whether to authenticate or not. The intermediate server will have no ability to view or interfere with the exchange.

\subsubsection*{Future Directions}

\paragraph{\textit{Aliases}} \textsc{ssh} servers need to allow client log-ins despite the client using any number of names for the server. Allowing multiple aliases for one server could be realized by supporting the newly standardized WebAuthn related origin requests (\textsc{ror}) in the implementation.

\paragraph{\textit{Tunneling}} Another requirement for a replacement is working anywhere traditional \textsc{ssh} authentication works. This presents an obstacle in using web technologies in this setting, as for example WebAuthn requires an \textsc{http} transport over ports 80 or 443, while ssh is traditionally on port 22. When these ports are not accessible, they may need to be tunneled through a \textsc{ssh} data channel. While this can be achieved with a flag on the client-side, this presents an unsatisfactory solution. Transparent, automatic tunneling would be preferable.

\paragraph{\textit{Attestation}} Another feature of the Web Authentication protocol of interest is \textit{attestation}, where the authenticator can be required to prove some aspect of the auth process. For example, to require biometric user verification or prove that the authenticator was manufactured by a specific trusted party. This offers distinct auditability advantages in an enterprise setting. However, we also note that risk of attestation being deployed in a manner that is user-hostile or anticompetitive.

\paragraph{\textit{Native support}} As mentioned previously, switching to a browser from the terminal to complete the authentication introduces friction. A tiered solution that maintains backwards compatibility might use native authentication if an ssh client with a ssh-passkey plugin is detected and fall-back to the browser based authentication for ssh clients without it.

\section{Related Works}

A centralized \textsc{ssh} key distribution method, relying on a trusted server, is proposed in~\cite{peng-10-device-identity}. Using the \textsc{dns} infrastructure to disseminate \textsc{ssh} host key (but not user) fingerprints has also been proposed, bringing \textsc{pki}-like authentication to \textsc{ssh}~\cite{schlyter-06-dns-ssh-fingerprint}. \textit{Perspectives} suggests a method of verifying host keys by collaborating with ``notary'' hosts~\cite{wendlandt-08-perspectives}. Another proposal recommends using the \textit{Tor} onion routing network to perform secondary key verification~\cite{alicherry-09-multi-path-verify}. One approach to reducing user-key vulnerability to theft is using \textsc{rsa} threshold signatures to distribute the key among user devices, such that more than one compromise is required for a key leak~\cite{harchol-18-dist-ssh-key}. Proposals to improve the privacy properties of key-based ssh authentication have been proposed to avoid leaking all the client's public keys to their counterparty~\cite{roy-22-pracical-pets}. These approaches are not adopted in practice~\cite{herley-09-so-long}.
 
\textsc{Ssh} certificates are used by systems like Netflix's \textsc{bless}~\cite{netflix-16-bless}. Amazon Web Services offers \textsc{ec2} \textit{Instance Connect}, where \textsc{ssh} keys can be pushed to hosts automatically based on user permissions---each user must still manage their own keys~\cite{aws-20-connect-linux}.

A new protocol, \textit{Remote Terminals over HTTP/3} (previously called \textsc{Ssh}3 but incompatible with the SSH protocol), has been proposed to incorporate some web technologies such as \textsc{quic} and \textsc{tls} 1.3 into the remote server administration~\cite{michel-24-remote-terminal}. This proposal supports relying on third parties for authentication using the \textsc{o}Auth protocol~\cite{hardt-12-oauth-framework}.

It appears that \textsc{ssh} keys are the classic user rejection of security mechanisms due to costly externalities~\cite{herley-09-so-long}. Well-known poor usability of \textsc{ssh} keys. Users do not verify fingerprints~\cite{gutmann-11-users-verify-keys}. They do not accurately self-report security practices either~\cite{wash-17-self-report}.

\textsc{Nist} released a standard to specify best practices for \textsc{ssh} key management~\cite{ylonen-15-sec-interactive}. Survey work has summarized the desiderata in \textsc{ssh} key management approaches and identifies it as an important open problem~\cite{ylonen-19-challenges}.

A few prior works on passkeys: one considers whether the \textsc{Fido}2 standard will finally be the ``kingslayer'' of passwords~\cite{lyastani-20-kingslayer}. This study finds that users are willing to accept passkeys as a replacement for passwords. However, they report apprehensions with the possibility of theft: without a mechanism for authenticator revocation, users are worried about the potential for account compromise following theft. Some users also reported discomfort with the physical form factor. Earlier studies on the \textsc{fido}2 precursor, \textsc{u}2\textsc{f}, as a second-factor report good overall usability, with $2\times$ reduction in login times for average users compared to other 2-factor authentication methods~\cite{lang-16-practical-scnd-factor}.

\section{Conclusion}
We propose the utilization of passkeys for \textsc{ssh} authentication, with the \textsc{ssh}-passkeys framework. We develop this framework to integrate support in a backwards- and client-compatible by utilizing pluggable authentication modules (\textsc{pam}) and the Web Authentication \textsc{api}. We develop a prototype. We evaluate the prototype formally, comparing it with 36 other authentication schemes, and find favorable security, deployability and usability characteristics. We experimentally validate the prototype in a user study, confirming the poor usability of key-based baselines and the ability of our approach to remedy those known problems. We find our approach has superior security and usability. We discuss the implications of our findings in context of the current security landscape and the remaining steps for realization of this approach.

\ifCLASSOPTIONcompsoc
  \section*{Acknowledgments}
\else
  \section*{Acknowledgment}
\fi

Moe Kayali was partially supported by NSF IIS 2314527 and NSF SHF 2312195. We gratefully acknowledge funding support from the Brett Helsel Career Development Professorship.

% \IEEEtriggeratref{26}



\bibliographystyle{IEEEtran}
\bibliography{bibliography}

\begin{thebibliography}{10}
\providecommand{\url}[1]{#1}
\csname url@samestyle\endcsname
\providecommand{\newblock}{\relax}
\providecommand{\bibinfo}[2]{#2}
\providecommand{\BIBentrySTDinterwordspacing}{\spaceskip=0pt\relax}
\providecommand{\BIBentryALTinterwordstretchfactor}{4}
\providecommand{\BIBentryALTinterwordspacing}{\spaceskip=\fontdimen2\font plus
\BIBentryALTinterwordstretchfactor\fontdimen3\font minus \fontdimen4\font\relax}
\providecommand{\BIBforeignlanguage}[2]{{%
\expandafter\ifx\csname l@#1\endcsname\relax
\typeout{** WARNING: IEEEtran.bst: No hyphenation pattern has been}%
\typeout{** loaded for the language `#1'. Using the pattern for}%
\typeout{** the default language instead.}%
\else
\language=\csname l@#1\endcsname
\fi
#2}}
\providecommand{\BIBdecl}{\relax}
\BIBdecl

\bibitem{ylonen-06-ssh-prot}
\BIBentryALTinterwordspacing
T.~Ylonen and C.~M. Lonvick, ``{The Secure Shell (SSH) Protocol Architecture},'' Internet Requests for Comments, {RFC Editor}, {RFC} 4251, Jan. 2006. [Online]. Available: \url{https://tools.ietf.org/html/rfc4251}
\BIBentrySTDinterwordspacing

\bibitem{ylonen-15-sec-interactive}
T.~Ylonen, P.~Turner, K.~Scarfone, and M.~Souppaya, ``{Security of Interactive and Automated Access Management Using Secure Shell (SSH)},'' Oct. 2015.

\bibitem{alata-06-lessons-learned}
E.~Alata, V.~Nicomette, M.~Ka{\^{a}}niche, M.~Dacier, and M.~Herrb, ``{Lessons Learned from the Deployment of a High-Interaction Honeypot},'' in \emph{Sixth European Dependable Computing Conference ({EDCC})}, ser. EDCC~'06.\hskip 1em plus 0.5em minus 0.4em\relax Coimbra, Portugal: {IEEE} Computer Society, Oct. 2006, pp. 39--46.

\bibitem{andrews-20-prevalence-pass}
R.~Andrews, D.~A. Hahn, and A.~G. Bardas, ``{Measuring the Prevalence of the Password Authentication Vulnerability in SSH},'' in \emph{International Conference on Communications}, ser. ICC~'20.\hskip 1em plus 0.5em minus 0.4em\relax Dublin, Ireland: {IEEE} Computer Society, Jun. 2020, pp. 1--7.

\bibitem{ylonen-19-challenges}
T.~Ylonen, ``{Challenges in Managing SSH Keys -- and a Call for Solutions},'' University of Helsinki, Tech. Rep., 2019.

\bibitem{w3c-21-webauthn}
\BIBentryALTinterwordspacing
{W3C Web Authentication Working Group}, ``{Web Authentication: An API for accessing Public Key Credentials Level 3},'' {World Wide Web Consortium (W3C)}, {W3C Recommendation}, Apr. 2021. [Online]. Available: \url{https://www.w3.org/TR/webauthn/}
\BIBentrySTDinterwordspacing

\bibitem{caniuse-25-webauthn}
{Can I use}, ``{Web Authentication API},'' 2025, \url{https://caniuse.com/webauthn}, as of \today.

\bibitem{bonneau-12-the-quest}
J.~Bonneau, C.~Herley, P.~C. Van~Oorschot, and F.~Stajano, ``{The Quest to Replace Passwords: A Framework for Comparative Evaluation of Web Authentication Schemes},'' in \emph{IEEE Symposium on Security and Privacy}, ser. SP~'12.\hskip 1em plus 0.5em minus 0.4em\relax San Jose, California, USA: {IEEE} Computer Society, May 2012, pp. 553--567.

\bibitem{galbraith-06-ssh-key-file}
\BIBentryALTinterwordspacing
J.~Galbraith and R.~L. Thayer, ``{The Secure Shell (SSH) Public Key File Format},'' Internet Requests for Comments, {RFC Editor}, {RFC} 4716, Nov. 2006. [Online]. Available: \url{https://datatracker.ietf.org/doc/html/rfc4716}
\BIBentrySTDinterwordspacing

\bibitem{lerner-17-confidante}
A.~Lerner, E.~Zeng, and F.~Roesner, ``{Confidante: Usable Encrypted Email: A Case Study with Lawyers and Journalists},'' in \emph{IEEE European Symposium on Security and Privacy}, ser. EuroS{\&}P~'17.\hskip 1em plus 0.5em minus 0.4em\relax Paris, France: {IEEE} Computer Society, Apr. 2017, pp. 385--400.

\bibitem{whitten-99-johnny-encrypt}
A.~Whitten and J.~D. Tygar, ``{Why Johnny Can't Encrypt: A Usability Evaluation of PGP 5.0},'' in \emph{USENIX Security Symposium}, ser. SSYM~'99.\hskip 1em plus 0.5em minus 0.4em\relax Washington, District of Columbia, USA: USENIX, Aug. 1999, pp. 169--183.

\bibitem{oasis-22-pkcs}
\BIBentryALTinterwordspacing
O.~Open, ``{PKCS \#11 Specification Version 3.1},'' OASIS, Committee Specification Draft~01, Feb. 2022. [Online]. Available: \url{https://docs.oasis-open.org/pkcs11/pkcs11-spec/v3.1/csd01/pkcs11-spec-v3.1-csd01.html}
\BIBentrySTDinterwordspacing

\bibitem{igoe-11-cert-ssh}
\BIBentryALTinterwordspacing
K.~Igoe and D.~Stebila, ``{X.509v3 Certificates for Secure Shell Authentication},'' Internet Requests for Comments, {RFC Editor}, {RFC} 6187, Mar. 2011. [Online]. Available: \url{https://datatracker.ietf.org/doc/html/rfc6187}
\BIBentrySTDinterwordspacing

\bibitem{openSSH-20-release}
{OpenSSH Development Team}, ``{OpenSSH Release Notes 8.2},'' Feb. 2020, \url{https://www.openssh.com/txt/release-8.2}, as of \today.

\bibitem{kovacs-14-reactions-sony}
E.~Kovacs, ``{Industry Reactions to Devastating Sony Hack},'' Dec. 2014, \url{https://www.securityweek.com/industry-reactions-devastating-sony-hack/}, as of \today.

\bibitem{sony-14-20f}
{Sony Corporation}, ``{Form 20-F Annual Report for the Fiscal Year Ended March 31, 2014},'' {U.S. Securities and Exchange Commission Filing}, 2014, \url{https://www.sony.com/en/SonyInfo/IR/library/FY2014_20F_PDF.pdf}, as of \today.

\bibitem{foster-15-fast-and-vuln}
I.~D. Foster, A.~Prudhomme, K.~Koscher, and S.~Savage, ``{Fast and Vulnerable: A Story of Telematic Failures},'' in \emph{USENIX Workshop on Offensive Technologies}, ser. WOOT~'15.\hskip 1em plus 0.5em minus 0.4em\relax Washington, District of Columbia, USA: USENIX, Aug. 2015.

\bibitem{lang-16-practical-scnd-factor}
J.~Lang, A.~Czeskis, D.~Balfanz, M.~Schilder, and S.~Srinivas, ``{Security Keys: Practical Cryptographic Second Factors for the Modern Web},'' in \emph{Financial Cryptography and Data Security}, ser. Lecture Notes in Computer Science, vol. 9603.\hskip 1em plus 0.5em minus 0.4em\relax Christ Church, Barbados: Springer, Feb. 2016, pp. 422--440.

\bibitem{samar-95-pam-unified}
V.~Samar and R.~Schemers, ``{Unified Login with Pluggable Authentication Modules (PAM)},'' Sun Corporation, Request For Comments, Oct. 1995.

\bibitem{boyd-96-essence}
J.~R. Boyd, ``{The Essence of Winning and Losing},'' Air University Library, Jan. 1996, \url{https://fasttransients.files.wordpress.com/2010/03/essence_of_winning_losing.pdf}, as of \today.

\bibitem{dell-12-yours-is-better}
N.~Dell, V.~Vaidyanathan, I.~Medhi, E.~Cutrell, and W.~Thies, ``{``Yours is better!'': participant response bias in HCI},'' in \emph{ACM Conference on Human Factors in Computing Systems}, ser. CHI~'12.\hskip 1em plus 0.5em minus 0.4em\relax Austin, Texas, USA: ACM, May 2012, pp. 1321--1330.

\bibitem{brooke-86-sus}
J.~Brooke, ``{System Usability Scale (SUS)},'' Agency for Healthcare Research and Quality (AHRQ), Tech. Rep., 1986, \url{https://digital.ahrq.gov/sites/default/files/docs/survey/systemusabilityscale%2528sus%2529_comp%255B1%255D.pdf}, as of \today.

\bibitem{dhs-12-menlo}
{Department of Homeland Security}, ``{The Menlo Report: Ethical Principles Guiding Information and Communication Technology Research},'' U.S. Department of Homeland Security, Science and Technology Directorate, Tech. Rep., Aug. 2012, \url{https://www.dhs.gov/sites/default/files/publications/CSD-MenloPrinciplesCORE-20120803_1.pdf}, as of \today.

\bibitem{roy-22-pracical-pets}
L.~Roy, S.~Lyakhov, Y.~Jang, and M.~Rosulek, ``{Practical Privacy-Preserving Authentication for SSH},'' in \emph{USENIX Security Symposium}, ser. SSYM~'22.\hskip 1em plus 0.5em minus 0.4em\relax Boston, Massachusetts, USA: USENIX, Aug. 2022, pp. 3345--3362.

\bibitem{valsorda-20-whoami}
F.~Valsorda, ``{whoami.filippo.io: A SSH Server that knows who you are},'' GitHub repository, 2020, \url{https://github.com/FiloSottile/whoami.filippo.io}, as of \today.

\bibitem{wendlandt-08-perspectives}
D.~Wendlandt, D.~G. Andersen, and A.~Perrig, ``{Perspectives: Improving SSH-style Host Authentication with Multi-Path Probing},'' in \emph{USENIX Annual Technical Conference}, ser. ATC~'08.\hskip 1em plus 0.5em minus 0.4em\relax Boston, Massachusetts, USA: USENIX, Jun. 2008, pp. 321--334.

\bibitem{rivest-84-expose-eave}
R.~L. Rivest and A.~Shamir, ``{How to Expose an Eavesdropper},'' \emph{Communications of the ACM}, vol.~27, no.~4, pp. 393--395, Apr. 1984.

\bibitem{wash-17-self-report}
R.~Wash, E.~J. Rader, and C.~Fennell, ``{Can People Self-Report Security Accurately? Agreement Between Self-Report and Behavioral Measures},'' in \emph{ACM Conference on Human Factors in Computing Systems}, ser. CHI~'17.\hskip 1em plus 0.5em minus 0.4em\relax Denver, Colorado, USA: ACM, May 2017, pp. 2228--2232.

\bibitem{florencio-07-web-pass}
D.~Flor{\^{e}}ncio and C.~Herley, ``{A Large-Scale Study of Web Password Habits},'' in \emph{The World Wide Web Conference}, ser. WWW~'07.\hskip 1em plus 0.5em minus 0.4em\relax Banff, Alberta, Canada: {ACM}, May 2007, pp. 657--666.

\bibitem{bangor-24-empirical-sus}
A.~Bangor, P.~T. Kortum, and J.~T. Miller, ``{An Empirical Evaluation of the System Usability Scale},'' \emph{International Journal of Human--Computer Interaction}, vol.~24, no.~6, pp. 574--594, Jul. 2008.

\bibitem{peng-10-device-identity}
J.~Peng and X.~Zhao, ``{SSH-Based Device Identity and Trust Initialization},'' \emph{Information Security Journal: A Global Perspective}, vol.~19, no.~5, pp. 237--242, Oct. 2010.

\bibitem{schlyter-06-dns-ssh-fingerprint}
\BIBentryALTinterwordspacing
J.~Schlyter and W.~Griffin, ``{Using DNS to Securely Publish Secure Shell (SSH) Key Fingerprints},'' Internet Requests for Comments, pp. 1--9, Jan. 2006. [Online]. Available: \url{https://datatracker.ietf.org/doc/html/rfc4255}
\BIBentrySTDinterwordspacing

\bibitem{alicherry-09-multi-path-verify}
M.~Alicherry and A.~D. Keromytis, ``{Doublecheck: Multi-Path Verification Against Man-in-the-Middle Attacks},'' in \emph{Symposium on Computers and Communications}, ser. ISCC~'09.\hskip 1em plus 0.5em minus 0.4em\relax Sousse, Tunisia: {IEEE} Computer Society, Jul. 2009, pp. 557--563.

\bibitem{harchol-18-dist-ssh-key}
Y.~Harchol, I.~Abraham, and B.~Pinkas, ``{Distributed SSH Key Management with Proactive RSA Threshold Signatures},'' in \emph{Applied Cryptography and Network Security}, ser. ACNS~'18, vol. 10892.\hskip 1em plus 0.5em minus 0.4em\relax Leuven, Belgium: Springer, Jul. 2018, pp. 22--43.

\bibitem{herley-09-so-long}
C.~Herley, ``{So Long, and No Thanks for the Externalities: The Rational Rejection of Security Advice by Users},'' in \emph{New Security Paradigms Workshop}, ser. NSPW~'09.\hskip 1em plus 0.5em minus 0.4em\relax Oxford, United Kingdom: {ACM}, Sep. 2009, pp. 133--144.

\bibitem{netflix-16-bless}
I.~Netflix, ``{BLESS: An SSH Certificate Authority that runs as an AWS Lambda function},'' May 2016, \url{https://github.com/Netflix/bless}, as of \today.

\bibitem{aws-20-connect-linux}
{Amazon Web Services}, ``{Connect to a Linux instance using EC2 Instance Connect},'' 2020, \url{https://docs.aws.amazon.com/AWSEC2/latest/UserGuide/ec2-instance-connect-methods.html}, as of \today.

\bibitem{michel-24-remote-terminal}
\BIBentryALTinterwordspacing
F.~Michel and O.~Bonaventure, ``{Remote terminal over HTTP/3 connections},'' Internet Engineering Task Force, Internet-Draft draft-michel-remote-terminal-http3-00, Aug. 2024. [Online]. Available: \url{https://datatracker.ietf.org/doc/draft-michel-remote-terminal-http3/00/}
\BIBentrySTDinterwordspacing

\bibitem{hardt-12-oauth-framework}
\BIBentryALTinterwordspacing
D.~Hardt, ``{The OAuth 2.0 Authorization Framework},'' Internet Requests for Comments, {RFC Editor}, {RFC} 6749, Oct. 2012. [Online]. Available: \url{https://datatracker.ietf.org/doc/html/rfc6749}
\BIBentrySTDinterwordspacing

\bibitem{gutmann-11-users-verify-keys}
P.~Gutmann, ``{Do Users Verify SSH Keys?}'' \emph{;login:}, vol.~36, no.~4, pp. 35--36, Aug. 2011.

\bibitem{lyastani-20-kingslayer}
S.~G. Lyastani, M.~Schilling, M.~Neumayr, M.~Backes, and S.~Bugiel, ``{Is FIDO2 the Kingslayer of User Authentication? A Comparative Usability Study of FIDO2 Passwordless Authentication},'' in \emph{IEEE Symposium on Security and Privacy}, ser. SP~'20.\hskip 1em plus 0.5em minus 0.4em\relax Virtual Conference: IEEE, May 2020, pp. 268--285.

\end{thebibliography}

\end{document}